\documentclass[12pt]{iopart}

\usepackage{iopams}  
\usepackage{graphicx}
\usepackage{subfig}
\usepackage{units}
\usepackage{url}

\begin{document}

\title{Production of Neutral Pions and Eta-mesons in pp Collisions Measured with ALICE}

\author{Klaus Reygers, for the ALICE collaboration}

\address{University of Heidelberg,  Physikalisches Institut, Philosophenweg 12,\\69120 Heidelberg, Germany}
\ead{reygers@physi.uni-heidelberg.de}
\begin{abstract}
Invariant cross sections for neutral pions and $\eta$ mesons in pp collisions at $\sqrt{s} = 0.9$, 2.76, and  \unit[7]{TeV} were measured by the ALICE detector at the Large Hadron Collider. Next-to-leading order (NLO) perturbative QCD calculations describe the $\pi^0$ and $\eta$ spectra at \unit[0.9]{TeV}, but overestimate the measured cross sections at \unit[2.76]{TeV} and \unit[7]{TeV}. The measured $\eta/\pi^0$ ratio is consistent with $m_T$ scaling at \unit[2.76]{TeV}. At \unit[7]{TeV} indications for a violation of $m_T$ scaling were found.
\end{abstract}


\section{Introduction}
Transverse momentum ($p_T$) spectra of neutral pions and $\eta$ mesons in pp collisions at the LHC are interesting for several reasons. First, the measurement of these spectra helps characterize particle production in a new energy regime and allows to test phenomenological rules like $m_T$ scaling and $x_T$ scaling observed at lower $\sqrt{s}$. More importantly, these measurements test perturbative QCD at the highest available energies. With the current data in the range $p_T < \unit[30]{GeV}/c$ one constrains the gluon fragmentation function used in these calculations. Moreover, $\pi^0$ and $\eta$ spectra in pp are a basic ingredient in the extraction of direct photons as they are the dominant sources of the photons from hadron decays. Finally, $\pi^0$ and $\eta$ $p_T$ spectra in pp collisions provide a reference for the study of nuclear effects in Pb-Pb collisions with the aid of the nuclear modification factor \cite{Balbastre:2011}.

Perturbative QCD calculations are based on a factorization of phenomenological parton distribution and fragmentation functions (FF) and calculable parton cross sections. The parton distribution functions in the Bjorken-$x$ range around $x \approx 0.01$, relevant for the data presented here, are well under control. Gluon fragmentation functions, however,  are not so well constrained from  $e^+e^-$ data as gluon jets are a sub-leading NLO correction in these reactions \cite{deFlorian:2007aj}. Pion and $\eta$ meson production at moderate $p_T \lesssim \unit[30]{GeV}/c$ at the LHC, on the other hand,  is dominated by gluon fragmentation. For instance, at $\sqrt{s} = \unit[7]{TeV}$ about 90\,\% of the neutral pions at $p_T = \unit[10]{GeV}/c$ come from gluon fragmentation \cite{vogelsang}. At the RHIC energy of $\sqrt{s} = \unit[0.2]{TeV}$ the fraction of neutral pions from gluon fragmentation at the same $p_T$ is about $40\,\%$ \cite{vogelsang}. Hence, the measurement of $\pi^0$ and $\eta$ spectra in collisions of hadrons provides important constraints for the gluon fragmentation function.

Gluon fragmentation functions from different authors are available for the use in perturbative QCD calculations, e.g., KRE\,\cite{Kretzer:2000yf}, BKK\,\cite{Binnewies:1994ju}, KKP\,\cite{Kniehl:2000fe}, AKK\,\cite{Albino:2005me}, DSS\,\cite{deFlorian:2007aj}. Neutral pion spectra measured in pp collisions at $\sqrt{s} = \unit[0.2]{TeV}$ at RHIC favor gluon fragmentation functions with large average pion multiplicities (KKP, AKK and DSS are favored over KRE) \cite{deFlorian:2007aj}.

The ALICE detector~\cite{Aamodt:2008zz} at the LHC measures photons and neutral mesons in three different ways: with PHOS, with EMCAL, and with photon conversions.  PHOS is a homogenous electromagnetic calorimeter made of lead tungstate (PbWO$_4$) crystals. It subtends a pseudo-rapidity range $|\eta| < 0.13$ and an azimuthal range $\Delta \phi = 60^\circ$. EMCAL is a lead scintillator sandwich calorimeter. In its current form it subtends $|\eta| < 0.7$ and $\Delta \phi = 100^\circ$ ($\Delta \phi = 40^\circ$ in 2010). Electron-Positron pairs from photon conversions between the beam pipe and the middle of the Time Projection Chamber (TPC) (radial distance $R \approx \unit[180]{cm}$) can be detected with the TPC \cite{Koch:2011fw,Aamodt:2011}. The detector material in this range ($11.4\,\% $ of a radiation length $X_0$) corresponds to a photon conversion probability of $p_\mathrm{conv} \approx 8.5\,\%$. The lower $p_T$ threshold for the detection of an electron or positron is $p_{T,\mathrm{min}} \approx \unit[50]{MeV}/c$. The measurement of photon conversions provides a detailed study of the material budget of the ALICE experiment. Up to the middle of the TPC the ALICE material budget agrees within $+3.4\,\%/-6.2\,\%$ with its implementation in the GEANT simulation package. 

\section{Results}
\begin{figure}
\centering
\subfloat[]{\includegraphics[width=0.5\linewidth]{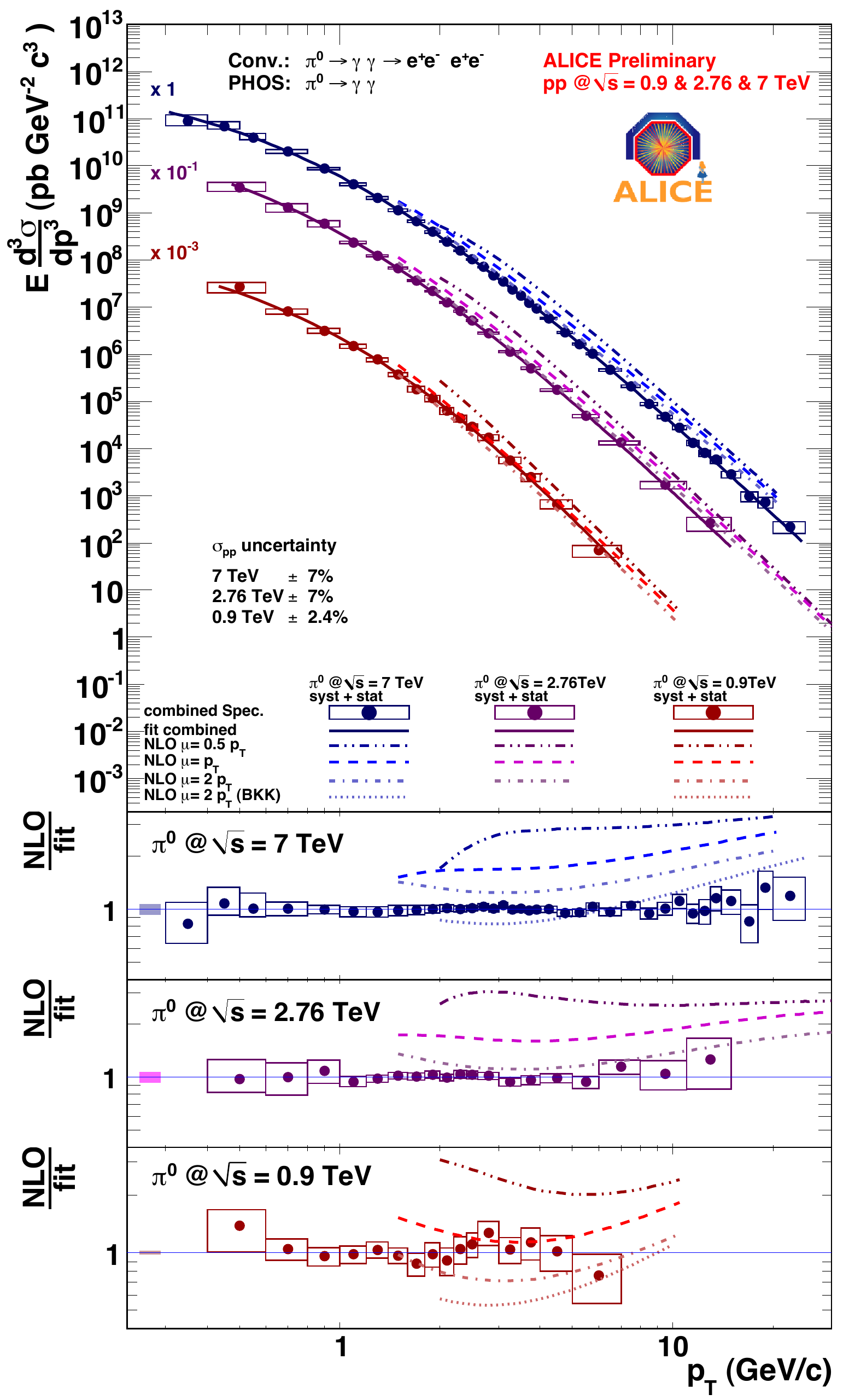}}
\subfloat[]{\includegraphics[width=0.5\linewidth]{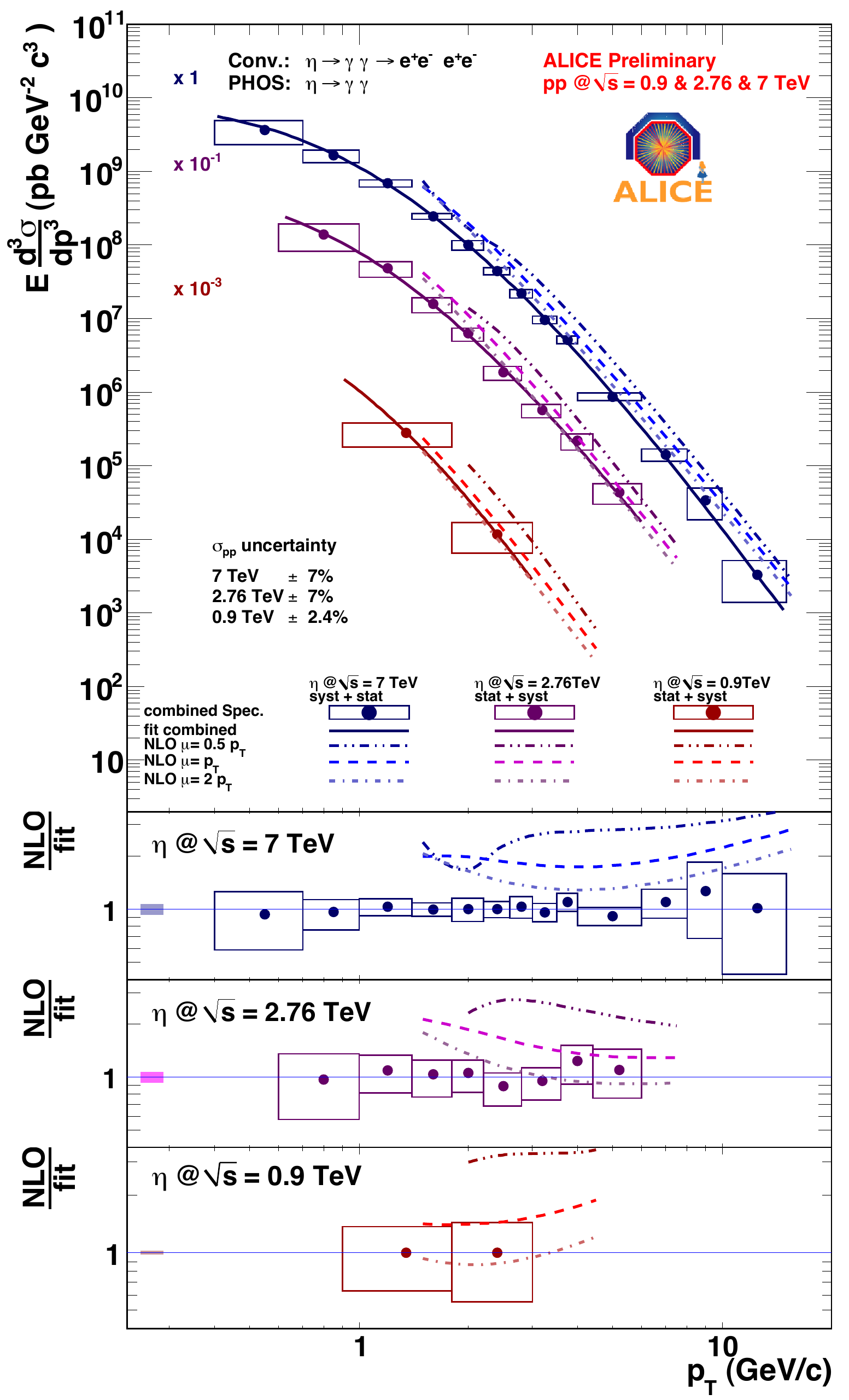}}
\caption{Invariant cross sections for neutral pions (a) and $\eta$ mesons (b) in pp collisions at $\sqrt{s} = 0.9$, 2.76, and  \unit[7]{TeV}. The spectra represent the combined result from the conversion method and from PHOS. The data were parameterized with a Tsallis function $E\,\mathrm{d}^3\sigma/\mathrm{d}p^3 \propto (1 + (m_T -m)/(nT))^{-n}$ and compared with a next-to-leading order perturbative QCD calculation.}
\label{fig:pi0_eta_cross_section}
\end{figure}

The three different methods to measure photons and neutral mesons in ALICE have different systematics and therefore provide valuable cross checks. In this paper results from PHOS and the conversion method are presented. The results of these two methods agree. The combined invariant cross sections from PHOS and the conversion method at $\sqrt{s} = 0.9$, 2.76, and  \unit[7]{TeV} are shown in Figure~\ref{fig:pi0_eta_cross_section} for $p+p \rightarrow \pi^0 +X$ (a) and $p+p \rightarrow \eta +X$ (b). The spectra are fully corrected, including a small (at maximum 20\,\%) correction of the yield such that the cross section given by a data point reflects the true cross section at the $p_T$ of the point, independent of the bin width \cite{Lafferty:1994cj}. 

The measured $\pi^0$ and $\eta$ cross sections are compared to NLO pQCD calculations \cite{Aversa:1988vb, Jager:2002xm} which use the CTEQ6M5  parton distribution functions \cite{Pumplin:2002vw} and the DSS  fragmentation functions \cite{deFlorian:2007aj}. The theoretical uncertainty of the calculation was estimated by varying the unphysical renormalization ($\mu_R$), factorization ($\mu_F$), and fragmentation scales ($\mu_{F'}$): $\mu = \mu_R = \mu_F = \mu_{F'} = 0.5 p_T, p_T, 2 p_T$. The NLO pQCD calculation with the DSS FF describes the $\pi^0$ spectrum at $\sqrt{s} = \unit[0.9]{TeV}$, but overestimates the data at $\sqrt{s} = \unit[2.76]{TeV}$ and $\sqrt{s} = \unit[7]{TeV}$ for all choices of the scales. The BKK FF \cite{Binnewies:1994ju}, used here in conjunction with a INCNLO calculation \cite{incnlo}, shows a better agreement with the data $\sqrt{s} = \unit[7]{TeV}$. For the $\eta$ cross section the NLO calculation was done with the AESSS FF \cite{Aidala:2010bn}. A similar trend as for the $\pi^0$ is found: the NLO pQCD agrees with the measured $\eta$ cross sections at $\sqrt{s} = \unit[0.9]{TeV}$, whereas the calculation overestimates the data at $\sqrt{s} = \unit[7]{TeV}$.

\begin{figure}
\centering
\subfloat[]{\includegraphics[width=0.49\linewidth]{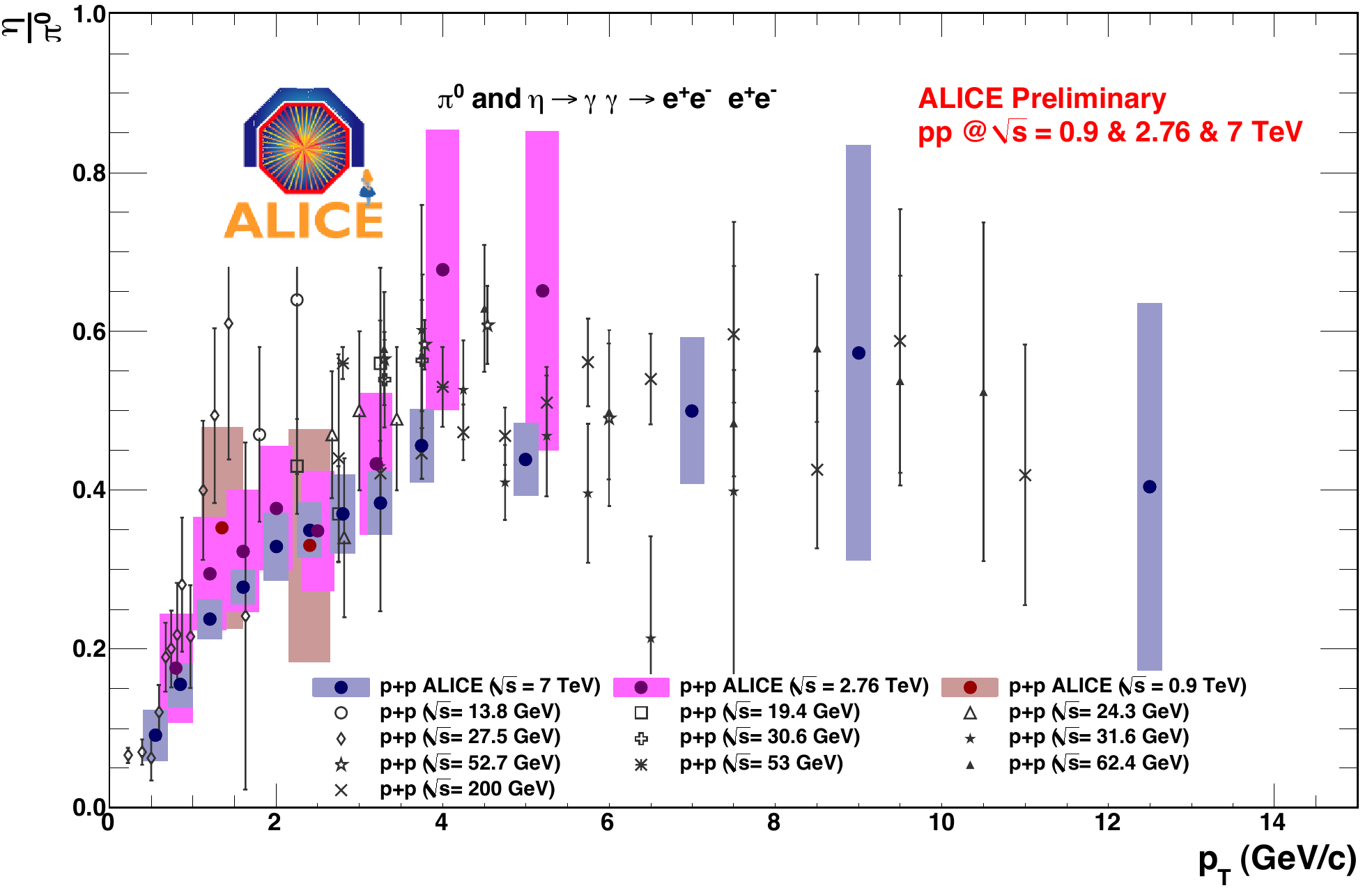}}
\subfloat[]{\includegraphics[width=0.51\linewidth]{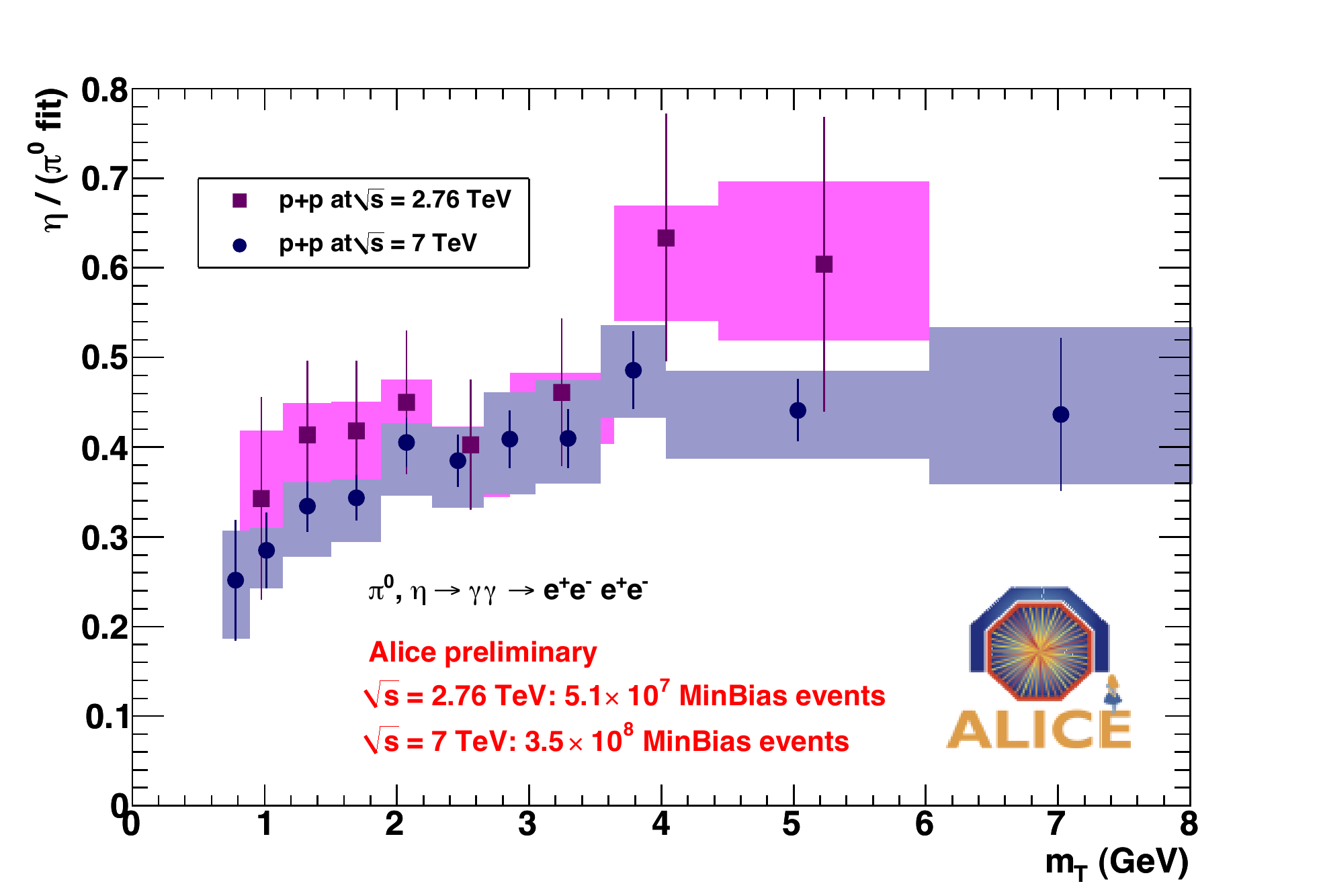}}
\caption{a) $\eta/\pi^0$ ratio as measured by ALICE in pp collisions at $\sqrt{s} = 0.9$, 2.76, and  \unit[7]{TeV} compared to the world's data. b) $\eta/\pi^0$ ratio at $\sqrt{s} = 2.76$, and  \unit[7]{TeV} as a function of transverse mass $m_T = \sqrt{p_T^2 + m^2}$.}
\label{fig:eta_pi0_ratio}
\end{figure}

The $\eta/\pi^0$ ratios in pp collisions at $\sqrt{s} = 0.9$, 2.76, and \unit[7]{TeV} as measured by ALICE are compared to the world's data for pp($\bar \mathrm{p}$) collisions at lower $\sqrt{s}$ in Figure~\ref{fig:eta_pi0_ratio}a. The asymptotic value of $\eta/\pi^0 \approx 0.4 - 0.6$ at high $p_T$ as well as the rise at low $p_T$ at LHC energies follows the trend observed at lower $\sqrt{s}$. In order to test $m_T$ scaling the $\eta/\pi^0$ ratio is plotted as a function of $m_T$ in Figure~\ref{fig:eta_pi0_ratio}b. The $\eta/\pi^0$ ratio at $\sqrt{s} = \unit[2.76]{TeV}$  is consistent with being constant (i.e., consistent with $m_T$ scaling) whereas at $\sqrt{s} = \unit[7]{TeV}$ there appears to be a violation of $m_T$ scaling.

\section{Conclusions}
Neutral pion and $\eta$ invariant cross sections were measured in pp collisions at $\sqrt{s} = 0.9$, 2.76, and \unit[7]{TeV}. A NLO pQCD calculation using the  CTEQ6M5 parton distribution functions and the DSS fragmentation functions describes the $\pi^0$ and $\eta$ spectra at \unit[0.9]{TeV} and overestimates the data at \unit[2.76]{TeV} and \unit[7]{TeV}. The $\eta/\pi^0$ ratio was studied at all three energies. An indication for $m_T$ scaling violation was found at $\sqrt{s} = \unit[7]{TeV}$.

\vspace{0.3cm}
\noindent{\it Acknowledgments}:
We would like to thank Werner Vogelsang for providing the NLO QCD calculations used in this paper.

\end{document}